\RequirePackage{fix-cm}
\documentclass[smallextended]{svjour3}       
\smartqed  
\usepackage{graphicx}

\begin{document}

\title{X(3872) revisited: 
the roles of OPEP 
and the quark degrees of freedom
}


\author{Sachiko Takeuchi\and Yasuhiro Yamaguchi\and Atsushi Hosaka  \and
        Makoto Takizawa 
}


\institute{S.\ Takeuchi \at
Japan College of Social Work, Kiyose, Tokyo 204-8555, Japan\\ 
\email{s.takeuchi@jcsw.ac.jp}           
\and
Y.\ Yamaguchi \at
Advanced Science Research Center, JAEA, 
Tokai, Ibaraki 319-1195, Japan
\and
A.\ Hosaka \at
Research  Center  for  Nuclear  Physics  (RCNP),  Ibaraki,  Osaka  567-0047,  Japan
\and 
M.\ Takizawa \at
Showa Pharmaceutical University,  Machida,  Tokyo  194-8543,  Japan
}

\date{Received: date / Accepted: date}

\maketitle

\begin{abstract}
The $X(3872)$ is investigated by employing the quark-hadron hybrid model,
that consists of the $c\bar c$ core,
$D^{(*)}\bar D{}^*$, $J/\psi\omega$, and $J/\psi\rho$ two-meson states.
Due to the attraction from the 
$c\bar c$-$D\bar D{}^*$ coupling and from the OPEP tensor coupling, 
a very thin peak can appear at the $D^{0}\bar D{}^{*0}$ threshold.
The energy of the corresponding pole of the scattering matrix is
$E=+0.06-0.14i$ MeV, which is on the physical sheet and above the threshold,
the same as the one of the poles from the LHCb data analysis.

\keywords{X(3872) \and exotic hadrons \and the quark model \and $S$-matrix poles}
 \PACS{
14.40.Rt 
 \and 
12.39.Jh 
}
\end{abstract}

\section{Introduction}
\label{intro}
The $X$(3872) is the well-known exotic hadron candidate, which was first found by Belle experiment 
in the weak decay of the $B$ meson,
$B^{\pm} \to J/\psi \, \pi^+ \, \pi^- \, K^{\pm}$ \cite{Choi:2003ue}.
It was soon confirmed 
by many types of the experiments:
the $B^{\pm,0}$ decay, the $p \bar p$ or $pp$ collision, and the $e^+ e^- \to \gamma X(3872)$ reaction
\cite{Zyla:2020zbs,Yamaguchi:2019vea}.
The $X(3872)$ decays into several kinds of the final states by the strong interaction: 
$D\bar D\pi$, $J/\psi \pi\pi$,  $J/\psi \pi\pi\pi$, and $\chi_{c1}(1P) \pi$.
Also, the production rate suggests that it contains the $c\bar c$ component 
by about five percent\cite{Yamaguchi:2019vea}.
The observations of the $X(3872)$ by various types of experiments strongly suggest 
that the $X(3872)$ is a resonance (or a bound state) rather 
than a kinematical peak enhanced, {\it e.g.}, by the triangle singularity. 
On the other hand, the mass of the $X(3872)$ peak observed in the final $J/\psi \pi\pi$ mass spectrum
corresponds 
to the $D^0\bar D{}^{*0}$ threshold within the experimental errors.
It is natural to consider that the peak is a cusp, or a virtual state of these two mesons.

Recent data analysis on the $X(3872)$ by LHCb
\cite{Aaij:2020qga}
shows that the $S$-matrix of this system has
 two poles, II and III, nearby the $D^0\bar D{}^{*0}$ threshold, whose central values are
$E_{\rm II} = +0.06-0.13i$ MeV and 
$E_{\rm III} = -3.58-1.22i$ MeV.
The pole II is on the physical sheet but its position is above the $D^0\bar D{}^{*0}$ threshold.
This does not correspond to a simple two-particle resonance nor a virtual state.

Whether the $X(3872)$ is a resonance, a cusp, or a bound state is still under discussion.
However, the observed sharp peak is a clear sign of the attraction 
that is strong enough to make the structure at the threshold.
Many types of explanations, including the meson-molecular-$c\bar c$ hybrid models 
\cite{Takizawa:2012hy,Takeuchi:2014rsa,Ferretti:2013faa}, have been proposed.
The $S$-matrix pole analysis by using the Fratt\'e approximation was also discussed in \cite{Hanhart:2011jz}.

In \cite{Takizawa:2012hy,Takeuchi:2014rsa}, the $X(3872)$
was investigated by a quark-hadron hybrid model, 
which is essentially a hadron model, 
but the interaction is constructed in accordance with the quark degrees of freedom. 
The $X$(3872) is a very shallow bound state or a virtual state in that model, 
which consists of the $D\bar D{}^*$ molecular states 
and a small amount of the $c\bar c$ core. 
In that picture, the attraction to form the peak comes from the $D\bar D{}^*$-$c\bar c$ coupling.
The spectra of the $c\bar c(2P)$ state decaying into the $D\bar D{}^*$ as well as the $\rho J/\psi$ and $\omega J/\psi$ are calculated by the model.
Both of the final $\rho J/\psi$ and $\omega J/\psi$ spectra were found to have a thin peak at  the $D^0\bar D{}^{*0}$ threshold. 

Since the picture that the $X(3872)$ is the two-meson molecule with the $c\bar c$ core
seems to explain many of the observed features of $X(3872)$,
we further expand the picture to include the one pion exchange (OPEP) between the $D$ and $\bar D{}^*$ mesons,
which surely exists.
In this proceedings, we briefly discuss how the hybrid model works well to explain the structure of X(3872).

\section{Model}
\label{sec:1}

In this work, the $X(3872)$ is investigated by employing the quark-hadron hybrid model used in refs. \cite{Yamaguchi:2019vea,Takizawa:2012hy,Takeuchi:2014rsa}.
Here we newly introduce the OPEP between $D^{(*)}$ and $\bar D{}^*$ mesons.
The model consists of  $c\bar c$,
$D^{(*)}\bar D{}^*$, $J/\psi\omega$, and $J/\psi\rho$.
We take only the $\chi_{c1}(2P)$ as the $c\bar c$ core, whose mass is
the closest to the $X(3872)$ mass.
The $D^{(*)}\bar D{}^*$ now includes $D\bar D{}^*(^3S_1)$, $D\bar D{}^*(^3D_1)$, and $D^*\bar D{}^*(^5D_1)$
due to the OPEP tensor coupling.
There is no isospin symmetry breaking term in the Hamiltonian except for the meson mass difference between
$D^{(*)0}$ and $D^{(*)\pm}$ and between the $\rho$ and the $\omega$ meson masses and widths.

The Hamiltonian of the whole system ${\cal H}$ can be expressed as 
\begin{eqnarray}
{\cal H}&=&
\left(\begin{array}{cc}
H&W\\
W^\dag&E^{(0)}_{c\bar c}
\end{array}\right)
,~~~~
H=
 H_0 + V + V_{\rm OPEP}
\end{eqnarray}%
where $H$ stands for the Hamiltonian within the two meson space,
$E^{(0)}_{c\bar c}$ the $1^{++}$ $c\bar c$ mass without the coupling to the two-meson states.
$W$ stands for the transfer potential between the two-meson states and the $c\bar c$ core,
and $V$  for the potential within the two-meson channels: 
\begin{eqnarray}
W(q)&=&{w_i\over \sqrt{\Lambda_q}}{\Lambda_q^2\over \Lambda_q^2+q^2}
,~~~~
V(q)={v_{ij}\over \Lambda_q^2}{\Lambda_q^2\over \Lambda_q^2+q^2}{\Lambda_q^2\over \Lambda_q^2+{q'}^2}
\\[2ex]
w_{i}&=&
\left(\begin{array}{r}
-g_0   \\
g_0 \\
0~   \\
0~    \\
\end{array}
\right),~~~
v_{ij}=
\left(\begin{array}{rrrrr}
  v & 0 & -u & -u \\
 0 & v& u & -u\\
 -u  & u &v' & 0\,\\
 -u & -u &0\, & v' \\
\end{array}
\right) ,
\label{eq:X-UV}
\end{eqnarray}%
for the $S$-wave $D{}^0 \bar D{}^{*0}$, 
$D^\pm D^{*\mp}$, $J/\psi\omega$ and $J/\psi\rho$ channels, respectively.

The model here essentially follows the parameter set A in ref.\ \cite{Takeuchi:2014rsa}.
We assume that $E^{(0)}_{c\bar c}$ to be the $\chi_{c1}(2P)$ mass calculated by the quark model,
3950 MeV.
The transfer potential $W$ exists only between the $c\bar c$ and the $S$-wave $D\bar D{}^*$ channels.
Its coupling strength  $g_0$ is a free parameter in this model.
It is assumed that there is no direct coupling between
the $c\bar c$ and the $J/\psi\omega$ 
because of the OZI rule, or the $J/\psi\rho$  because of the isospin symmetry.
For the potential $V$, we omit the empirical central potential in ref.\ \cite{Takeuchi:2014rsa}
($v=v'=0$),
and vary the strength $u$ for the coupling between the $J/\psi\omega$($\rho$)
and the $D\bar D{}^*$ $S$-wave channels.
This coupling is considered to come from the quark rearrangement, whose size depends much on the quark model parameters.
Here we show the results with the one twice as large as the parameter set in ref.\ \cite{Takeuchi:2014rsa}.

As for the OPEP, we use the central and tensor terms as:
\begin{eqnarray}
V_{\rm OPEP}(\vec{r})&=&\sum_{i<j}
\left({g_\pi\over 2f_\pi}\right)^2{1\over 3}[\vec{S}_i\cdot\vec{S}_jC(r,\mu)+\tens{S}_{12}T(r,\mu)]\tau_i\cdot\tau_j,
\end{eqnarray}%
where $C$ and $T$ are the Yukawa central and tensor terms with the Lorentzian cutoff $\Lambda_\pi$, respectively \cite{Yamaguchi:2019vea}.
We do not include the contact term in the central part.
From the quark model viewpoint, 
suppose there is an interaction between the $q$ and $\bar q$ in the $D^{(*)}\bar D{}^*$ channel,
the same interaction should also work between the $q$ and $\bar q$ within the $\omega$ or $\rho$ mesons.
Since the observed $\omega$ and $\rho$ meson mass difference is small,
the short range part of the isospin dependent interaction is considered to be weak.
So, here we introduce only the long range Yukawa term for the OPEP.

The energy spectrum of the $c\bar c$ decay to anything is proportional to the
imaginary part of the full propagator. With the self energy of the $c\bar c$ state, $\Sigma_{c\bar c}$, it can be written as \cite{Takizawa:2012hy,Takeuchi:2014rsa}
\begin{eqnarray}
\sum_f {dW(c\bar c\to f)\over dE} &=& -{1\over \pi} Im \langle c\bar c|{1\over E-E^{(0)}_{c\bar c}-\Sigma_{c\bar c}}|c\bar c\rangle
\label{eq:5}
\end{eqnarray}%

\section{Results}
\label{sec:2}

In Table \ref{tab:1}, we list the parameters in the present model.
In the model from \underline{a} to \underline{d}, we change the $c\bar c$-$D\bar D{}^*$ coupling $g_0$
 so that the binding energy varies.
The system is solved by using the complex scaling method \cite{Myo:2014ypa}.
We have found two scattering poles in each of the parameter sets.
One of the poles corresponds to a broad resonance originated from the $\chi_{c1}(2P)$
with an energy of $\sim$3950 MeV. 
The other pole appears close to the $D^{0}\bar D{}^{*0}$ threshold.
The energy of the latter pole, $E$, 
is listed in Table \ref{tab:1}
together with the one calculated without the $\rho$ or $\omega$ meson width, $E'$.
The one corresponds to the pole III in the LHCb analysis, 
if exists, cannot be seen by this complex scaling method
being below the threshold on the unphysical sheet.

\begin{table}
\caption{Model Parameters}
\label{tab:1}       
\begin{tabular}[t]{cccccccccc}
\hline\noalign{\smallskip}
$\Lambda_q$   & $u$ &  $g_\pi$ & $\Lambda_\pi$ \\
\noalign{\smallskip}\hline\noalign{\smallskip}
0.5 GeV &  0.386  &  0.55 & 1.13 GeV \\
\noalign{\smallskip}\hline
\end{tabular}
\bigskip

\begin{tabular}[t]{cccccccccc}
\hline\noalign{\smallskip}
 & $g_0$ & $E$ (MeV) & $E'$ (MeV)  \\
\noalign{\smallskip}\hline\noalign{\smallskip}
\underline{a} & 0.0371 & $-1.98-0.51i$ & $-2.20$\\
\underline{b} & 0.0343 & $-0.94-0.44i$ & $-1.20$\\
\underline{c} & 0.0310 & $-0.13-0.27i$ & $-0.41$\\
\underline{d} & 0.0295 & $+0.06-0.14i$ & $-0.20$\\
\multicolumn{2}{c}{LHCb (pole II)\cite{Aaij:2020qga}} & $+0.06-0.13i$ & -\\
\noalign{\smallskip}\hline
\end{tabular}
\end{table}

\begin{figure}

\hspace*{6mm}  \includegraphics[clip, bb=0 0 389 287,width=0.75\textwidth]{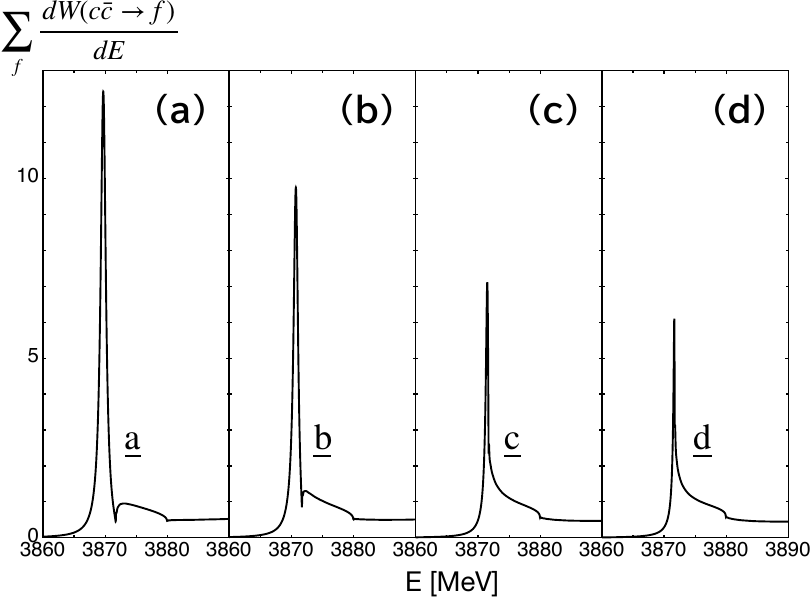}
  \bigskip
  
  \includegraphics[clip, bb=0 0 443 329,width=0.6\textwidth]{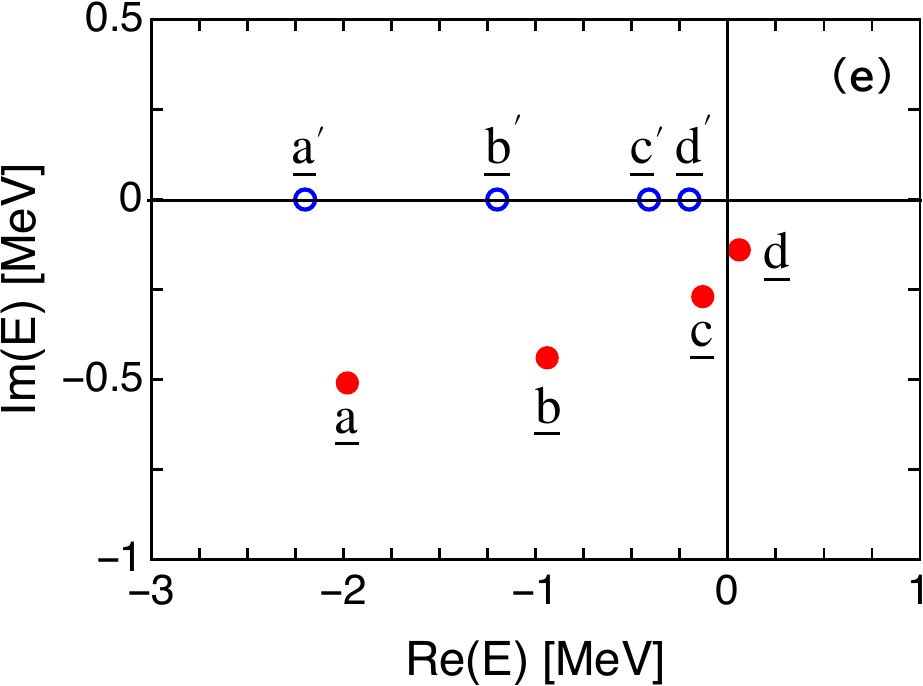}
\caption{The $c\bar c$ decay spectra to anything and poles of the scattering matrix.  
Figs.(a)-(d) show the decay spectrum with the parameter set \underline{a}-\underline{d}, respectively.
Fig.(e) shows the pole energy for each of the parameter sets.
}
\label{fig:1}       
\end{figure}

The energy spectra of the $c\bar c$ decay to anything
defined by Eq.\ (\ref{eq:5}) \cite{Takizawa:2012hy,Takeuchi:2014rsa} 
for the parameter set \underline{a}-\underline{d} are plotted in Fig.\ \ref{fig:1}(a)-(d), respectively.
In Fig.\ \ref{fig:1}(e), we plot the energies of the poles close  to the  $D^{0}\bar D{}^{*0}$ threshold.
The pole energy $E$ of each parameter set
is denoted as \underline{a}-\underline{d}, while $E'$ is denoted by \underline{a}$'$-\underline{d}$'$.
It is noteworthy that 
the pole \underline{d} appears on the physical sheet 
indicating that the pole is not a usual resonance.
In fact, without the constituting $\rho$ or $\omega$ meson width, 
the state corresponds to a bound state, \underline{d}$'$, with a binding energy of 0.20 MeV.
The pole energy of \underline{d} is nearly equals to the central value
of the LHCb data analysis, and that of \underline{c} is inside the 3$\sigma$ region \cite{Aaij:2020qga}.



The attraction that makes the $X$(3872) in this model 
comes both from the $c\bar c$-$D\bar D{}^*$ coupling and from the OPEP tensor term. 
By introducing the OPEP the coupling strength $g_0^2$ 
to make $E'=-0.20$ MeV reduces from $(0.0390)^2$ to $(0.0295)^2$ by 0.57 times.
The OPEP contributes by about half of the attraction in this parameter set 
and the model still successfully gives the $X(3872)$ peak.

\begin{acknowledgements}
This work is supported in part by 
JSPS KAKENHI Grant Numbers JP16K05361 (S.T. and M.T.), JP20K14478 (Y.Y.), 
 JP17K05441, Grants-in-Aid for Scientific Research on Innovative Areas (No.\ 18H05407) (A.H.),
and RCNP-CORENet.
\end{acknowledgements}


\end{document}